
\input harvmac.tex
\hfuzz 15pt



\def\za{\alpha} \def\zb{\beta} \def\zg{\gamma} \def\zd{\delta}

 \def\zp{\pi} \def\zr{\rho} 
 
\def\21{\alpha_{2\, 1}}\def\zgg{\gamma '}

\def\ba{a^*}

\def\zG{\Gamma}

\def\[{\,[\!\!\![\,} \def\]{\,]\!\!\!]\,}
\def\dC{C\kern-6.5pt I}

\def\L{\bar{L}}

\def\hhz_1{{\hat z}_1}
\def\hbz_2{{\hat {\bar z}}_2}

\def\ta{\tilde{\alpha}}
\def\tb{\tilde{\beta}}
\def\tg{\tilde{\gamma}}\def\tgg{\tilde{\gamma}'}

\def\CA{{\cal A}}              
\def\CD{{\cal D}}

%

%

%

\def\({ \left( }
\def\){ \right) }

\def\dal{
\vbox{
\halign to5pt{\strut##&
\hfil ## \hfil \cr
&$\kern -0.5pt
\sqcap$ \cr
\noalign{\kern -5pt
\hrule}
}}\ }

\def\dim{{\rm dim\,}}\def\mod{{\rm mod\,}}

\catcode`\@=11
\def\Eqalign#1{\null\,\vcenter{\openup\jot\m@th\ialign{
\strut\hfil$\displaystyle{##}$&$\displaystyle{{}##}$\hfil
&&\qquad\strut\hfil$\displaystyle{##}$&$\displaystyle{{}##}$
\hfil\crcr#1\crcr}}\,}   \catcode`\@=12



\def\IC{\relax\hbox{$\inbar\kern-.3em{\rm C}$}}

\input amssym.def
\input amssym.tex
\def\IR{\Bbb R}\def\IC{\Bbb C}


\def\UNIT{\blacktriangleleft\kern-.6em\blacktriangleright}

\def\la{\langle} \def\ra{\rangle}

\def\endPROOF{\quad$\blacksquare$\medskip}


\lref\BPZ{A.A. Belavin, A.M. Polyakov and A.M. Zamolodchikov,
Nucl. Phys. B241 (1984) 333.}

\lref\DF{Vl.S. Dotsenko and V.A. Fateev,
Nucl. Phys. B240 [FS 12]
 (1984) 312.}

\lref\FP{ P. Furlan and V.B. Petkova, Mod. Phys. Lett.  A4
(1989) 227.}

\lref\SY{ K. Symanzik, Lettere al Nuovo Cimento, 3 (1972) 734.}

\lref\AKF{ P. Appell,  J. Kamp\'e de F\'eriet: {\it Fonctions
Hyperg\'eom\'etriques et
Hypersph\'eriques - Polynomes
 d' Hermite}, Gauthier - Villars, Paris, 1926.}
\lref\Bat{ H. Bateman: {\it Higher Transcendental Functions}, vol.
I, McGraw-Hill, New York, 1953.}

\lref\DS{D.T. Stoyanov,
 Infinite dimensional Lie algebras connected with the four-dimensional
Laplace operator, ISAS preprint 10/85/E.P., and
in the  Proceedings of the  {\it Symposium on
Conformal Groups and Structures}, Clausthal, 1985, Eds. A.O.
Barut and H.D. Doebner, Lecture Notes in Physics, 
261 (1986) 
379,
 Springer-Verlag, Berlin.}

\lref\GR{I.S. Gradshteyn and I.M. Ryzhik,
{\it Table of Integrals, Series and Products}, Academic Press, New York,
1980.}

\lref\ITT{I.T. Todorov, Infinite dimensional Lie algebras in
conformal QFT models,
in the  Proceedings of the  {\it Symposium on
Conformal Groups and Structures}, Clausthal, 1985, Eds. A.O.
Barut and H.D. Doebner, Lecture Notes in Physics,
261, (1986)
 387,
Springer-Verlag, Berlin.}

\lref\F{G. Felder,
Nucl. Phys. B317 (1989) 215.}

\lref\NST{N. M. Nikolov, Ya. S. Stanev and I.T. Todorov,
     J.Phys. A35 (2002) 2985,
hep-th/0110230.}

\lref\NSTb{N.M. Nikolov, Ya. S. Stanev and I.T. Todorov, 
       Nucl.Phys. B670 (2003) 373, hep-th/0305200.}


%

%

\Title{\vbox{\baselineskip12pt\hbox
{  }\hbox{  }}}
{\vbox{\centerline
 {Virasoro Symmetry in  a 2h-dimensional Model    }
\medskip
\centerline{ and Its
Implications}
 \vskip2pt
}}
 
\bigskip\bigskip

\centerline{P. Furlan$^{*, \dagger}\, $ and  $\ $  V.B. Petkova$^{**}$}
\bigskip
\bigskip
\vbox{\centerline{\it $^{*}$Dipartimento  di Fisica Teorica
 dell'Universit\`{a} di Trieste, Italy,}
\medskip
\centerline{\it $^{\dagger}$Istituto Nazionale di Fisica Nucleare (INFN),
Sezione di Trieste, Italy,}
\medskip
\centerline{\it $^{**}$ Institute for Nuclear Research and
Nuclear Energy, Sofia,
Bulgaria}
}
\medskip

\bigskip\bigskip

\vskip 1.5cm

\centerline{\bf Abstract}
\bigskip
\medskip

 \noindent 
{The  set of two partial differential equations for the Appell
hypergeometric function in two variables
$F_4(\alpha ,\beta ,\gamma , \alpha+\beta-\gamma+2-h ; x,y)$
is shown to arise as a null  vector decoupling relation in a
2h-dimensional generalisation of the Coulomb gas model. It
corresponds to a level two  singular vector of an intrinsic
Virasoro  algebra.}
 
 \vskip 1.5cm
\medskip
\rightline{{\it
Dedicated to the memory of Mitko Stoyanov}}

\medskip
\medskip
\Date{ }
\vfil\eject

\baselineskip=14pt plus 1pt minus 1pt

%
%
\newsec{Introduction}

The hypergeometric function is an ubiquitous object of the
two-dimensional conformal field theories, providing examples of
4-point correlation functions of  various models.  The reason behind this
is that it  is the simplest example of a solution of null vector
decoupling equations associated with singular vectors in
Virasoro algebra Verma modules.  Thus the second order
hypergeometric equation appears   as a differential operator
realisation of a singular vector at level two  \BPZ.

Our aim in this paper is to demonstrate that a hidden Virasoro
algebra plays a similar role in a higher dimensional conformal
model. In particular the singular vector at level two gives rise
in even 2h-dimensional space-time to a pair of second order
linear partial differential equations. These are the Appell -
Kamp\'e de F\'eriet (AK) equations, \AKF, \Bat\
\eqna\part
$$\eqalignno{
&\Big(x(1-x)\,{\partial^2 \over \partial x^2} - y^2{\partial^2
\over \partial y^2}  -
2xy\,{\partial^2 \over \partial x\partial y}  +
\zg {\partial \over \partial x}
-(\za +\zb +1)(x {\partial \over \partial x} +
y {\partial \over \partial y}) -\za\zb \Big) F  = 0 \cr
& {} &  \part {}\cr
&\Big(y(1-y){\partial^2 \over \partial y^2} - x^2\,{\partial^2
\over \partial x^2}
-2xy\,{\partial^2 \over \partial x\partial y}  +
\zgg {\partial \over \partial y}
-(\za +\zb +1)(x {\partial \over \partial x} +y {\partial \over \partial y})
 -\za\zb \Big) F  = 0
}$$
 satisfied  by the Appell  hypergeometric functions of type
\eqn\F{
F_4(\za , \zb , \zg , \zgg ; x,y) =
\sum_{m,n = 0}^{+\infty}{(\za )_{m+n}(\zb )_{m+n}\over (\zg )_m(\zgg )_nm!n!}
x^my^n\,,
}
$  (\za)_{n}=\Gamma(\za+n)/\Gamma(\za)$,
with
\eqn\ddil{\za +\zb -\zg -\zgg = h-2\,.}
 The  two variables
 \eqn\anh{
x= {r_{13}r_{24}\over r_{14}r_{23}}\,,
\quad
y= {r_{12}r_{34}\over r_{14}r_{23}}\,; \qquad r_{ij} :=\, x_{ij}^2\,,
}
are the two anharmonic ratios, made of the coordinates $x_i\in \IR^{(2h)}$
 of a 4-point conformal invariant.
The model  is a 2h-dimensional
 generalisation of the two-dimensional
  Coulomb gas model with a charge at infinity \DF,
  described by  a (sub)-canonical
 field with logarithmic propagator,
 \eqn\prop{
 \la \phi(x_1) \phi(x_2)\ra \sim ((-\dal)^h)^{-1}=
-{1\over (4\pi)^h\,\Gamma(h)}\,\log x^2_{12}\,,
}
and scalar fields realised by vertex operators
$V_{\za}(x)=e^{i\za \phi(x)}$; it was studied in  \FP{}.  In the
two-dimensional case the  system of equations \part{}\ reduces, after
proper change of variables, to a linear combination of  two
(chiral)  hypergeometric equations.

\medskip

The appearance of a  Virasoro algebra in a four-dimensional
context was pointed out many years ago by Dimitar Stoyanov \DS,
who was studying the  infinite dimensional Lie algebras
preserving the solutions of the Laplace equation; one of the two
algebras he had constructed, contains  a subalgebra isomorphic to
the Virasoro algebra; see also
\NST, for a more recent development.

\subsec{The 4-point function}

We recall here briefly the construction  in \FP.

Consider the 4-point function in dimension $2h$ described by
only one integration.  It is written in terms of vertex operators
(VO) as
\eqna\vert
$$\eqalignno{
&\la \psi_4(x_4)\psi_3(x_3)\psi_2(x_2)\psi_1(x_1)\ra =
 \int {\rm d}^{2h}\,x_5\, \la V_{\za_+}(x_5)\, V_{\za_4}(x_4)\,
V_{\za_3}(x_3)\, V_{\za_2}(x_2)\, V_{\za_1}(x_1)\,
\ra \cr
&=\prod_{1\le i<j \le 4} \, r_{ij}^{2\za_i\za_j}\,
\int {\rm d}^{2h}x_5\, \prod_{i=1}^4\, r_{i5}^{-\delta_i}&\vert {}\cr
&=\pi^h \prod_i{1\over \Gamma(\delta_i)}\, \prod_{1\le i<j \le 4}
 r_{ij}^{2\za_i\za_j} \,  r_{12}^{h-\zd_1-\zd_2}\,
r_{13}^{h-\zd_1-\zd_3}\, r_{23}^{\zd_1-h}\,
r_{14}^{-\zd_4}\
F(x, y)\,, \quad \zd_i=-2 \za_+\, \za_i\,
}$$
and the  conformal invariance imposes the condition
\eqn\dil{
\sum_{i=1}^4\zd_i = 2h \ \  \leftrightarrow \ \
 \sum_{i=1}^4\,\alpha_i+ \alpha_+   = 2\alpha_0 \,
 (= \alpha_+ -{h\over \alpha_+}) \,. \qquad
}
The charges are parametrised as
\eqn\ala{
\za^J= J\, \sqrt{h\over t} = - J \alpha_+\,, \quad 2\za_0=
\sqrt{h}(\sqrt{t}-{1\over\sqrt{t}})\,, 
}
and the scaling dimension is
$d=2\triangle(\za)=2\za(\za-2\za_0)\,,$  or,
\eqn\dim{
\triangle(\za^J)=h\,\triangle_{J}\,, \
\,\triangle_{J}= J(J+1-t)/t=\triangle_{t-1-J}\,.
}
Following  \SY, $F(x,y)$
is given by the two fold Mellin  integral
\eqna\ISY
$$\eqalignno{
&
F(x,y)
 ={1\over (2\zp i)^{2}}\, \int_{\uparrow}{\rm d}s\,
\int_{\uparrow}{\rm d}t\,
x^{s}\, y^{t}\, \zG (-s)\, \zG (-t) \cr
&\zG (\zd_4+s+t)\, \zG (h-\zd_1+s+t)\, \zG
(\zd_1+\zd_2-h-t)\, \zG (\zd_1+\zd_3-h-s)
&\ISY {} \cr
}$$
with the paths of integration
running parallel to the imaginary axis.
Closing the
contours to the right
and taking into account the poles
of the gamma factors
produces a linear combination
 of four infinite sums, that  can be  identified with
 the four linearly independent solutions of the
AK equations
\eqn\gen{\eqalign{
F(x,y) = & A\, F_4(\za , \zb , \zg , \zgg ; x , y )  \cr
&+B \, x^{1-\zg}F_4 (\za +1-\zg ,\zb +1-\zg ,2-\zg , \zgg ; x, y )\cr
&+C\, y^{1-\zgg}F_4(\za +1-\zgg ,\zb +1-\zgg ,\zg , 2-\zgg ; x, y )\cr
&+D\, x^{1-\zg}y^{1-\zgg}F_4 (\za +2-\zg -\zgg ,\zb +2-\zg -\zgg ,2-\zg ,
2- \zgg ; x, y )\,.\cr
}}
Here
\eqn\grk{
\za = \zd_4\,, \ \ \zb = h-\zd_1\,, \ \ \zg = 1+h-\zd_1-\zd_3\,,  \ \
\zgg = 1+h-\zd_1-\zd_2\,,
}
and these parameters satisfy \ddil\ as a result of the constraint
\dil.

In the limit $y\to 0, x\to 1$ the four terms in \gen{}\ combine
to produce  two leading order terms in the powers of $y$. In \FP\
it was observed that when
$\za_1={1\over 2\sqrt{t}}$  these
two terms correspond to the contribution 
 of scalar fields of
dimensions $d=2 \triangle(\za_2 \pm \za_1)$. The same is true with the
 other choice of screening charge, i.e,
$\alpha_+ \to \, 2\alpha_0-\alpha_+$ and  $ \alpha_1 = -{\sqrt{t}\over 2}$.
  The  Symanzik method \SY\
was further used to analyse the small distance behaviour
in  the case of two integrals, i.e., two screening charges of any of the two 
types,
and respectively the choices
$\za_1={1\over  \sqrt{t}}\,, \za_1=-\sqrt{t}\,, \za_1=
{1\over 2}({1\over \sqrt{t}}-\sqrt{t})$. The result is  three
terms
in the first two cases, or four terms in the last
one, again in full analogy with the two-dimensional case. This
suggests that there is some hidden Virasoro symmetry in this
higher dimensional problem, which can explain in  algebraic terms
the results in \FP. 
In the next sections we shall develop an operator approach,
 which will allow to 
construct and exploit this algebra.

\newsec{Fock space quantisation of the 2h-dimensional sub-canonical field}

We choose complex Euclidean coordinates  $z_a= e^{i \tau}\, n_a\,, n\in
S^{2h-1}\subset \IR^{2h}\,, \tau\in\IC\,, $
$z^2=\sum_{a=1}^{2h} z_a^2$.
For real $\tau$  one recovers the compactified Minkowski space
$S^1 \times S^{2h-1}$.
We shall mostly use the  real Euclidean coordinates
$x_a=e^t\,n_a$ with
 $t=i\tau$ - real; both notations $z_a$ and $ x_a$ will
appear throughout the paper.
\medskip
The field $\phi(z)$ satisfying $\dal^h \phi(z) =(\sum_a
\partial_{z_a}^2)^h \, \phi(z) =
0$ admits
the  mode
  expansion
%
\eqn\dec{
\phi(z) = 2q - i\,b_0\, \log z^2+ 2i\, \sum_{n\ne 0}\,
{b_n(z)\over n} = 2q - i\, b_0\, \log z^2 +  2i\,
\sum_{n\ne 0}\,(z^2)^{-n\over 2}
\,  {b_n(\hat{z}))\over n}
}
with commutation relations
\eqna\com
$$\eqalignno{
&[b_n(z_1), b_{-m}(z_2)]=
  n \, \cos \, n \theta_{12}
\, \delta_{nm}\, ({z_2^2\over
z_1^2})^{n\over 2}\,,
\ \ \  [b_0, q]= - i\,,  &\com {}
}
$$
Here $\hat{z}=z/\sqrt{z^2}$,
$
\, \cos  \theta_{12}
=\hat{z_1}\cdot \hat{z_2}$ and  $\ b_n(\rho\, z)=\rho^{-n}\, b_n(z)$.
The one-dimensional projection  of \com{}\
with $z_i=\sqrt{z_i^2}\, e\,, e^2=1\,, 
\, i=1,2$ (so that $\cos \theta_{12}=1$)
reads
\eqn\coma{
[b_n(e), b_{-m}(e)]
= n\, \delta_{nm}\,.
}
It is assumed that
%
%
$\
b_n(z)\, |0\ra =0\,, 
\ \  \la 0|\, b_{-n}(z)=0\,, \ n \ge 0\,.$

\subsec{Relation to the  free field quantisation}

 For simplicity of presentation we restrict here to the
four-dimensional case, $2h=4$. The modes $b_n(z)\,, \, n\ne 0$
can be constructed  as linear combinations of the  free field
modes $a_n(z)$ described in  \ITT\
\eqn\frcom{
[a_n(z_1), a_{-m}(z_2)]= {1\over z_1^2}\, [\ba_{-n}({z_1\over
z_1^2}), a_{-m}(z_2)]=
 \delta_{nm}\,{1\over z_1^2}\, ({z_2^2\over z_1^2})^{n-1\over
2}\, C_{n-1}^{(1)}(\hat{z}_1\cdot \hat{z}_2) \,, n > 0
}
where $C_n^{(1)}(\cos \theta) ={\sin (n+1) \theta\over \sin
\theta} $.  The modes $a_n(z)$ are homogeneous $a_n(\rho\,
z)=\rho^{-n-1}\, a_n(z)$, harmonic variables $ \dal a_n(z) =0$.
For $n>0$, $a_{-n}(z)\,, \, \ba_{-n}(z)\,$ are polynomials,
realising an irrep of $SO(4)$ of  dim $n^2 $ (i.e., $a_{-n-1}(z)
= z_{\mu_1} ....z_{\mu_n}\, a_{\mu_1...\mu_n}$, where $
a_{\mu_1...\mu_n}$ are symmetric, traceless tensors), while
$a_n(z):= {1\over z^2}\, \ba_{-n}({z\over z ^2})$.

\medskip

Now we take two independent free fields, i.e.,
two commuting copies
$\{a_n\}\,, \, \{a'_n\}$,
$[a_n, a'_m]=0$, each set satisfying \frcom\ and define
\eqna\mod
$$\eqalignno{
b_{-n}(z) &= \sqrt{n\over 2}(
a_{-n-1}(z) + z^2\, a_{-n+1}^{'}(z))\,, \ n>0\cr
b_{n}(z) &=
 \sqrt{n\over 2}(z^2\, a_{n+1}(z) - \, a_{n-1}^{'}(z))
 \,,   \ n>0
&\mod {}
}
$$
so that
$$\dal^2 b_n(z)=0\,, \quad \ b_n(\rho\, z)=\rho^{-n}\, b_n(z)\,.
$$
Indeed \mod{}\ is the unique decomposition of the homogeneous
polynomial of degree $n$, subject to this equation, into a sum of
homogeneous harmonic polynomials. The  commutation relations
\frcom{}\ then imply \com{}.
The generalisation to $h > 2$ is straightforward with
the Gegenbauer polynomials $C_{n-h+1}^{(h-1)}(\cos \theta)$
 appearing in the r.h.s. of \frcom{}. 
In the  two-dimensional case $2h=2$ the free field modes
$b_{n}(z)$  split into a
sum of chiral pieces
\eqn\ac{
b_{-n}(z)={1\over 2}(e^{in (\tau+\sigma)}\, c_{-n}^{+} +e^{in
(\tau-\sigma)}\, c_{-n}^{-}) \,.
}

\subsec{Vertex operators}

 Let
\eqn\ver{
V_{\alpha}(z)=:e^{i\alpha \phi(z)}:=
(e^{i 2\alpha q}\,
 e^{i\alpha \phi_{<}(z)})\, ((z^2)^{\alpha\ b_0}\,
e^{i\alpha\phi_{>}(z)})
 =V_{\alpha}^-(z)\,\,V_{\alpha}^+(z)
}
where $ \phi_{{<\atop >}}(z)\,=\mp 2i\,\sum_{k>0}{b_{\mp k}(z)\over k}\,
$.  The commutation relations \com{}\ imply
\eqn\co{
[b_n(z_1), V_{\alpha}(z_2)]= 2\alpha\, ({z_2^2\over
z_1^2})^{n\over 2}\, \cos n \theta_{12}\,V_{\alpha}(z_2)\,.
}
Furthermore
using
the relation \GR\
\eqn\lg{
 \log ({z_{12}^2\over z_1^2})
 = \log(1-2\rho \,\cos  \theta + \rho^2)=- 2\sum_{n=1} \,\rho^n\,
{\cos \, n \theta \over n}
}
a standard  formula of the two-dimensional case is  generalised:
\eqn\op{
V_{\alpha_1}^+(z_1)\,V_{\alpha_2}^-(z_2)=\big(z_{12}^2\big)^{2\alpha_1\,
\alpha_2}\,
V_{\alpha_2}^-(z_2)\,V_{\alpha_1}^+(z_1)\,.
}
It implies the  operator product
expansion
\eqn\ope{
V_{\alpha_1}(z_1)\,V_{\alpha_2}(z_2)=\big(z_{12}^2\big)^{2\alpha_1\,
\alpha_2}\, V_{\alpha_1+\alpha_2}(z_2) + .....
}
 consistent with the 2-point function
\eqn\twop{
 \la 2\alpha_0|
 V_{2\alpha_0 -\alpha}(z_1)\,V_{\alpha}(z_2)|0\ra=
(z_{12}^2)^{-2\triangle(\alpha)}\,, \quad \triangle(\alpha)=
\za(\za-2\za_0)\,,
}
where $2\za_0$ parametrises the charge at infinity, i.e., we reproduce
\dim. 
The (normalised) bra and ket states are determined from the vertex operators
as
\eqna\brke
$$\eqalignno{
|\alpha \ra &=V_{\alpha}(0) |0\ra=e^{2i\, \alpha \, q}|0\ra\,,
& \brke {} \cr
\la \za|& =\la 0| e^{-2i\, \alpha  q}
=\lim_{x\to \infty} (x^2)^{2\triangle(\alpha)}\la 2\za_0|
\,  V_{2\alpha_0-\za}(x)\,.
}$$

\noindent
Having \ver\ one computes the matrix elements
\eqn\matre{
 \la 2\za_0- \za_{p+1} |V_{\za_p}(z_p) \dots V_{\za_2}(z_2) |\alpha_1\ra
=\prod_{1\le i<j\le p} (z^2_{ij})^{2\za_i\za_j}\,, \quad
\za_{p+1}=2\za_0-\sum_{i=1}^p
\za_i\,.
}The charge conservation condition in \matre\ implies the
 identities
\eqn\scal{
\sum_{i=1}^{p+1}\triangle(\za_i)=-\sum_{1 \le i<j\le p+1}
2\za_i\za_j\,\  \Longleftrightarrow
 \quad \sum_{i=1}^{p}\triangle(\za_i)-\triangle(\za_{p+1})
=-\sum_{1 \le i<j\le p} 2\za_i\za_j\,.
}
The integral of the VO with charge $\za=\za_+$, or $\za=2\za_0-\za_+\,,$
i.e., scaling dimension $2\triangle(\za_+)=2h$, provides the
2h-dimensional analog of the screening charge operator. In a
4-point matrix element with one screening charge we shall
use the notation $x_5,\za_5=\za_+$ keeping
the index $4$ for the last field in the 4-point function. Then
the matrix element is related to the $x_4\to \infty$ limit of the
4-point function in \vert{}\ according to
\eqna\exac
$$\eqalignno{
& \int d^{2h} x_5 \la \za_4|
V_{\za_+}(x_5) V_{\za_3}(x_3)
V_{\za_2}(x_2)|\za_1\ra = :\int  {\rm d}^{2h} x_5\   \CA \cr
&=(x_3^2)^{-\Delta}\,
x^{-2\alpha_2(\alpha_3-\za_1)}\,y^{2\alpha_{1}(2\alpha_0-\alpha_2)}\,
\
F(x,y)\,,
\ \  x={x_3^2\over x^2_{23}}\,, \  \ y={x_2^2\over x^2_{23}}\,, &
\exac {}
}$$
 where in agreement with \scal\ and using that $2\za_1=-\za_+$
 \eqn\powb{
\Delta:=\sum_{i=1}^3\, \triangle(\za_i)
- \triangle(2\za_0-\sum_{i=1}^3 \za_i
-\za_+) =2\za_1(\alpha_{2}+\alpha_{3}-2\za_0)-2\alpha_{2}\alpha_{3}
\,.}

\newsec{ A Virasoro algebra}

Analogously to the one-dimensional case  one can construct generators
\eqna\vir
$$\eqalignno{
\L_n(e)&= {1\over 2}\,
\sum_{k}\, b_{n-k}(e) \, b_{k}(e) -2\alpha_0(n+1)b_n(e)\,, \ n\ne 0\,,
\cr
\L_0(e)& = \sum_{n > 0}\, b_{-n}(e) \, b_{n}(e) +
{1\over 2}\, b_0^2 - 2\alpha_0 b_0 &\vir {} \,,
}$$
which  close, using the commutation relations \coma\ for collinear
vectors,
a Virasoro algebra  with central charge
\eqn\oldc{\eqalign{
&c=1-48 \alpha_0^2=1+ 24 h -12 h (t+{1\over t})\,, \cr
&\L_0 |\za^J \ra = 2\triangle(\za^J) |\za^J \ra = 2h\triangle_J |\za^J \ra\,.
}}
For $h=1/2$ one   recovers the
   conventional notation for the eigenvalues of these  two
   Virasoro generators. The algebra \vir{}\
  extends to $\L_{n}(z)=(\sqrt{z^2})^{-n}\, \L_{n}(\hat
z)$ (fixed $z$ in the commutator).

\medskip

Let us now  compute the action of the  generators \vir{}\
on the VO.  Using
\eqn\virf{
[b_k(\hat{z}_1), \L_{-n}(\hat{z}_2)] = \big(b_{k-n}(\hat{z}_2)\,
 + 2\za_0(n-1)\, \delta_{k,n}\big) k\,\cos k \theta_{12}\,,
}
we obtain, e.g. for $n>0$,  denoting $w:=\sqrt{z^2}$
\eqna\act
$$\eqalignno{
&[\L_{-n}(\hat{z}_1),V_{\alpha}(z_2)] = \cr
&2\alpha\,   w_2^{-n} \Big(\sum_{k> 0}\,
 w_2^{k}\, \cos (n-k) \theta_{12}\ b_{-k}(\hat{z}_1)\,
V_{\alpha}(z_2) +  \sum_{k\ge 0}\, w_2^{-k}\, \cos (n+ k)
\theta_{12}\,V_{\alpha}(z_2)\,
b_{k}(\hat{z}_1)\Big)  \cr
&
+2\alpha\, w_2^{-n} \Big(-2\alpha_0 (-n+1)   \, \cos n
\theta_{12}\, -\alpha \,  \sum_{k=1}^{n-1}\, \cos (n-
k) \theta_{12} \,  \cos k \theta_{12}\Big)
V_{\alpha}(z_2)\,    &\act {}  \cr
&=2\alpha\,\sum_{k}\,
 w_2^{k-n}\, \cos (n-k)\theta_{12}\
  :b_{-k}(\hat{z}_1) \,V_{\alpha}(z_2):
 -2\triangle(\alpha)\, (n-1)\, w_2^{-n}\,
  \cos n \theta_{12}\, V_{\alpha}(z_2)\cr
&+ \alpha^2\, w_2^{-n}
\big((n-1)\cos n \theta_{12}- {\sin (n-1)  \theta_{12} \over
 \sin \theta_{12} }\big)
 V_{\alpha}(z_2)\,.
}$$
The meaning of the normal product is
\eqn\norpr{
: b_k\, V:= b_k\, V\,, \  {\rm for}  \ k<0\,, \
\ : b_k\, V:=  V\, b_k\, \  {\rm for}  \ k \ge 0\,.
}
The second of the two  equalities in \act{}\  extends to
$\L_n\,, n\ge 0$ as well. In particular for $n=\pm 1, 0$ the very
last term in \act{}\ vanishes.

On a product of VO the (negative mode) generators act as
\eqna\acttb
$$\eqalignno{
&
[\L_{-n}(\hat{z}_1), V_{\alpha_p}(z_p)\,\dots
V_{\alpha_2}(z_2)]=\sum_{i=2}^p\, 2\alpha_i\,
\Big(2\alpha_0 (n-1) \,w_i^{-n}\, \cos n \theta_{1i}
V_{\alpha_p}(z_p)\, \dots V_{\alpha_2}(z_2)\cr
&+\sum_{k}\,  w_i^{k-n}\, \cos (n-k)\theta_{1i}\
:b_{-k}(\hat{z}_1) \big(V_{\alpha_p}(z_p)\,\dots \,V_{\alpha_2}(z_2)\big):\Big)
 & \acttb {} \cr
& -  2 \sum_{k=1}^{n-1}\,
\big(\sum_{i=2}^p w_i^{-n+k}\za_i\cos (n-k) \theta_{1i}\big) \,
\big(\sum_{j=2}^p w_j^{-k} \za_j\cos k \theta_{1j}\big)
\ V_{\alpha_p}(z_p)\, \dots  V_{\alpha_2}(z_2) \,, \cr
}$$
and
\eqn\actt{\eqalign{
&[\L_{0}(\hat{z}_1), V_{\alpha_p}(z_p)\, \dots
V_{\alpha_2}(z_2)]=2 \big(\sum_i \triangle(\za_i)+2 \sum_{i<j} \za_i\za_j \big)
V_{\alpha_p}(z_p)\,\dots  V_{\alpha_2}(z_2)\cr
&+\sum_{i=2}^p
\, 2\alpha_i\,\sum_{k}\,
 w_i^{k}\,
  \cos k\theta_{1i}\
  :b_{-k}(\hat{z}_1) \big(V_{\alpha_p}(z_p)\, \dots
\,V_{\alpha_2}(z_2)\big): \, \,.
}}
\medskip

In the  two-dimensional case  the expansion \dec\ splits
into chiral pieces due to the splitting \ac\ of the modes.
However the generators
 \vir{}\ do not reduce to a sum of chiral generators - there is a
mixed term surviving in this Virasoro algebra.

\subsec{Differential operators}

For $\hat{z}_2=\hat{z}_1$ \act{} reduces to a
differential operator
action 
\eqn\stv{
[\L_{n}(\hat{z}),V_{\alpha}(z)]= \Big(w^{n+1}\,
\partial_{w} +2\triangle(\alpha)\, (n+1)\,w^{n}\Big)
V_{\alpha}(z)\,.
}
This realisation by differential operators
 admits an extension to non-collinear arguments
e.g.,  for the matrix elements
\eqn\bil{\eqalign{
\la \za_4|[\L_{-n}(\hat{z_2}), V_{\alpha_3}(z_3)]\,
&V_{\alpha_2}(z_2)|\za_1\ra=D_{-n}(z_2,z_3)
\la \za_4| V_{\alpha_3}(z_3)\,V_{\alpha_2}(z_2)|\za_1\ra\,,\cr
D_{-n}(z_2,z_3)
&= w_3^{-n}\,\big(
{w_3\over w_2}{\sin n\theta\over
\sin \theta} \, z_2\cdot\partial_{z_3}-
{\sin (n-1)\theta\over \sin \theta} \,z_3\,\cdot\partial_{z_3}\big)\cr
 - & (2\triangle(\alpha_3)-\alpha_3^2)\, (n-1)\,   {\cos n
\theta_{23}\over w_3^{n}} -\alpha_3^2\, {\sin (n-1)  \theta_{23} \over
w_3^{n}\,  \sin \theta_{23} }
}}
These operators satisfy the  Witt algebra
\eqn\fdVirb{
[D_{-n},D_{-m}]
=(n-m)\, D_{-n-m}\,.
}

\subsec{Ward identity}
The  $\L_0$ Ward identity for the n-point correlation function coincides
with the 2h-dimensional dilatation Ward identity.
Indeed, we have
\eqna\derp
$$\eqalignno{
&w\partial_w\,V_{\alpha}(x)=x\cdot \partial_{x}\,V_{\alpha}(x)=
2\alpha\, \sum_{k}\,  :b_{-k}(x)\, V_{\alpha}(x):\,, & \derp {}\cr
&\big(\sum_i x_i\cdot\partial_{x_i}-\sum_{i<j} 4\za_i\za_j\big)\,
 V_{\alpha_p}(x_p)\,\dots V_{\alpha_1}(x_1) = \sum_i 2 \za_i \,
\sum_{k\ne 0 }\, :b_{-k}(x_i) \big(V_{\alpha_p}(x_p)\,\dots
\,V_{\alpha_1}(x_1)\big):
}$$
and
using the relations \scal,
\actt{},
one obtains for the matrix elements
\eqn\wid{\eqalign{
&0=\la \za_{p+1} |V_{\za_p}(x_p) \dots V_{\za_2}(x_2)
(2\triangle(\za_1)-\L_0(\hat{x}_0)) |\alpha_1\ra\cr
&=\big(\sum_{i=2}^{p} x_i\cdot\partial_{x_i}
+\sum_{i=1}^{p} 2 \triangle(\za_i) -2\triangle(\za_{p+1})\big)
\la  \za_{p+1} |V_{\za_p}(x_p) \dots V_{\za_2}(x_2) |\alpha_1\ra\,.
}}
The integrated version of \wid\ if, say,
$V_{\za_p}(x_p)$ is a screening charge operator,
i.e,  $\za_p=\za_+\,,\, 2\triangle(\za_+)=2h$, again reduces to the standard
Ward identity, obtained from \wid\ by skipping the last terms
in the two sums, since using that $  x_p\cdot\partial_{x_p}
+2h=\partial_{x_p} \cdot x_p$,
these two terms do not contribute to
the integral over $x_p$. In particular the Ward identity for the
matrix element in \exac{}\ reads
\eqn\ward{
\big(\sum_{i=2}^{3} x_i\cdot\partial_{x_i} +2\Delta\big) \int
d^{2h} x_5 \la \za_4| V_{\za_5}(x_5) V_{\za_3}(x_3)
V_{\za_2}(x_2)|\za_1\ra =0
}
which is also checked by the explicit expression in \exac{}\
using that $\sum_{i=2}^{3}\,(x_i\cdot\partial_{x_i})\, f(x,y)=0$
for any function  $f(x,y)$.

\subsec{Another realisation of the Virasoro algebra}

There is another
realisation of the generators of the Virasoro algebra
with the property that for $2h=1$ it again reproduces the known
one-dimensional  expressions,  namely
\eqna\newv
$$\eqalignno{
L_n(e)&= {1\over 2}\,
\sum_{k\ne 0,n}\, b_{n-k}(e) \, b_{k}(e) +{1\over \sqrt{2h}}\,
b_{n} b_0  -\sqrt{2\over h}\alpha_0(n+1)b_n(e)
\cr
& = \L_n(e) +({1\over \sqrt{2h}}-1)b_n(e)(b_0-2\alpha_0 (n+1))\,,
\ \ n\ne 0\,,  &\newv {}
\cr
L_0(e)& =
\sum_{n > 0}\, b_{-n}(e) \, b_{n}(e) +
{b_0\over 2h}({b_0\over 2} - 2\alpha_0)=
\L_0(e)+\big({1\over 2h}-1\big)b_0({b_0\over 2} - 2\alpha_0)\,.
}$$
As for  the first realisation in \vir{},
the duality transformation
$b_n \to 4\za_0 \, \delta_{n0}-b_{-n}\,$  sends
$L_n \to L_{-n}$, which is an automorphism of the Virasoro algebra.

More importantly, in
 this new  realisation  the eigenvalues of $L_0$ and the central charge
operator do not depend on $h$ and coincide with the  eigenvalues
in the one-dimensional case,
(cf. \ala{})
\eqn\standreal{\eqalign{
& c=1- {24 \over h}\alpha_0^2=13-6(t+{1\over t})\,,\cr
&L_0|\alpha^J\ra={\triangle(\alpha^J)\over
h}|\alpha^J\ra=\triangle_J|\alpha^J\ra \,.
}}
Hence in this realisation we can use all the standard expressions
for the singular vectors, as e.g.,
the singular vector at level two
\eqn\singv{
 \Big(t\,L_{-1}^2-L_{-2}\Big)\, |\za^{J={1\over 2}} \ra \,
}
and its counterpart with $t\to 1/t$ for $\za^{J=-{t\over 2}} $.
The new terms in \newv{}\ modify the  action \act{}\ of the generators  on the fields,
 and in particular the one-dimensional projection
 is no more  a simple
derivative with respect to w.
E.g. 
we get for the zero mode instead of \act{}\
\eqn\actb{\eqalign{
&[L_{0}(\hat{z}_1),V_{\alpha}(z_2)] =  \cr
& 2\za\,\sum_{k \ne 0}\, w_2^{k}\, \cos k \theta_{12}\ :b_{-k}(\hat{z}_1)
\,V_{\alpha}(z_2):   +{\triangle(\alpha) \over h}\,V_{\alpha}(z_2)  +2\za\,
({1\over 2h}-1)\, V_{\alpha}(z_2)\,   b_0 \,.
}}
Although the action of the zero mode
$L_0$ is modified, it does not spoil the validity of the
corresponding Ward identity.  Indeed, replacing in the first line
of \wid\ $\L_0 \to L_0\,, 2\triangle(\za_1) \to {\triangle(\za_1)\over
h}$, we recover again the second line, but multiplied with
$1/2h$.

\newsec{The null vector decoupling identities }

The true advantage of the new realisation \newv{}\ is confirmed by the
following
\medskip
{\bf Proposition:}
\medskip
\noindent
{\it Let $2\za_1=
\sqrt{h\over t}$
 so that
$\triangle(\za_1)=h({3\over 4t}-{1\over 2})$.
Then  the null vector decoupling identity}

\eqn\svid{
\la \za_{p+1} |V_{\za_p}(z_p) \dots V_{\za_2}(z_2) \big(t L_{-1}^2(\hat{z})-
L_{-2}(\hat{z})\big) |\alpha_1\ra
=0\,.
}
{\it holds true for any $z$.}
\medskip

The Proposition is proved by straightforward application of the
commutator  formulae derived above.
Applying $L_{-1}$ on the vector
$\la\za_{p+1},\dots,\za_2|:=
 \la\za_{p+1}| V_{\alpha_p}(z_p)\,\dots \,V_{\alpha_2}(z_2)
$

\noindent
we obtain
\eqna\acttbh
$$\eqalignno{
&-\la\za_{p+1}, \dots, \za_2 | L_{-1}(\hat{z}_0)= & \acttbh {}\cr
&\la\za_{p+1}, \dots, \za_2 | \sum_{i=2}^p
\, 2\alpha_i\,\Big(\sum_{k > 1}\, { \cos (1+k)\theta_{0i}\over w_i^{k+1}}
\,b_{k}(\hat{z}_0)  +  {\cos 2\theta_{0i}\over w_i^2}  b_1(\hat{z}_0)+
{1\over \sqrt{2h}}\,  {\cos \theta_{0i}\over w_i}  b_0\Big)\,.
}$$
Denoting
$\CA=\la \za_{p+1} |V_{\za_p}(z_p) \dots V_{\za_2}(z_2) |\alpha_1\ra
$
we have
\eqna\losq
$$\eqalignno{
-\la\za_{p+1},\dots, \za_2 |
\big(& L_{-1}( \hat{z}_0)\big) |\alpha_1\ra={1\over \sqrt{2h}}
\sum_{i=2}^p {4 \za_i \za_1\over w_i}\, \cos \theta_{i0}\,\CA
\,, \cr
\la\za_{p+1},\dots, \za_2 |& t\, L_{-1}^2( \hat{z}_0) |\alpha_1\ra =
& \losq {} \cr
&
\Big({t\over 2h}
\big(\sum_{i=2}^p 4 \za_i \za_1\, {\cos \theta_{i0}\over w_i}\big)^2
- {t\over \sqrt{2h}}\sum_{i=2}^p 4 \za_i \za_1\,
 {\cos 2\theta_{i0}\over w_i^2} \Big)\CA \,.
 }$$
Similarly using \acttb{}\ and \newv{}\ we compute
\eqna\lt
$$\eqalignno{
\la\za_{p+1}, \dots, \za_2 |&
\big(L_{-2}( \hat{z}_0)\big) |\alpha_1\ra =& \lt {}\cr
&\Big(2\big(\sum_{i=2}^p \za_i \, {\cos \theta_{i0}\over w_i}\big)^2
- {1\over \sqrt{2h}}\sum_{i=2}^p 4 \za_i (\za_1 +\za_0)\,
 {\cos 2\theta_{i0}\over w_i^2} \Big)\CA\,.
}$$
It remains to insert  the concrete value of $\za_1$ in \singv,
which implies
\eqn\valp{
(4\za_1)^2=(-2\za_+)^2={4h\over t}\,,
\qquad \za_1+\za_0
=\za_1\,t \,,
}
and hence \svid\  and the second equality in \losq{}\ coincide, thus proving \svid.\hskip -0.2cm \endPROOF

On the other hand we can partially express these matrix elements in terms
of differential operators.
Thus we get
\eqn\loh{\eqalign{
&-\la\za_{p+1}, \dots, \za_2 |
\big(L_{-1}(\hat{z}_0)\big) |\alpha_1\ra
={1\over \sqrt{2h}}
 \hat{z}_0\cdot
\big(\partial_{z_2}+ \dots +\partial_{z_p}\big) \, \CA
\,.
 }}
Furthermore we use
\eqn\direl{\eqalign{
&\la\za_{p+1}, \dots, \za_2 |\Big({\cos 2\theta_{0i}\over w_i^2}
b_1(\hat{z}_0) +{1\over \sqrt{2h}}\big(\hat{z}_0\cdot
\partial_{z_i} b_1(z_i)\big) \Big)
L_{-1}(\hat{z}_0)  |\za_1\ra=\cr
&{1\over \sqrt{2h}}{2\za_1\,\cos 2\theta_{0i}\over w_i^2}(1-{1\over \sqrt{2h}})
\CA\,.
}}
This equality is true also when $z_0$ coincides with some $z_i$
since then we can use the scaling property of the modes $(z\cdot
\partial_{z})\, b_n(z)=-n\, b_n(z)$.
Hence
\eqn\lohb{\eqalign{
&\la\za_{p+1}, \dots, \za_2 |L_{-1}^2(\hat{z}_0) |\alpha_1\ra
=\cr
&{1\over 2h} \big(\hat{z}_0\cdot
(\partial_{z_2}+ \dots +\partial_{z_p})\big)^2 \, \CA - \sum_{i=2}^p 4 \za_i
\za_1\,{\cos
2\theta_{0i}\over w_i^2}({1\over \sqrt{2h}}-{1\over 2h})\,
\CA\,.\cr
 }}
 Combining with \lt{}\ we can write
 \eqna\dinv
 $$\eqalignno{
&{2h\over t}
\la\za_{p+1}, \dots, \za_2 |\big(t L_{-1}^2(z_0)-L_{-2}(z_0)\big)
 |\alpha_1\ra
=\Big(\big(z_0\cdot
(\partial_{z_2}+ \dots +\partial_{z_p})\big)^2 & \dinv {}\cr
&-\sum_{i=2}^p \delta_{0i}\,
z_0\cdot(\partial_{z_2}+ \dots +\partial_{z_p})
-\big( \sum_{i=2}^p 4 \za_i \za_1\,{w_0\over w_i} \cos
\theta_{0i}\big)^2 +\sum_{i=2}^p 4 \za_i \za_1\,{w_0^2\over w_i^2}
\cos 2\theta_{0i}
\Big)\, \CA  \,.
}$$

\subsec{The 3-point  null vector decoupling condition}

We consider first the 3-point null vector decoupling condition,
which determines the possible "fusions" with the fundamental field
with $\triangle(\za^{J={1\over 2}})$.
The 3-point matrix element (with one screening charge)
is determined by the $L_0$ Ward identity as
\eqn\tps{\eqalign{
\int
\, \CA&=\int d^{2h} x_5 \,  \la \za_3|
V_{\za_+}(x_5) \,V_{\za_2}(x_2)|\za_1\ra=w_2^{2a}\,, \cr
-a&=\triangle(\za_1)+\triangle(\za_2)-\triangle(\za_3) \,.
}}
Choosing the argument of the Virasoro generators as  $z_0=x_2$
 we can represent \dinv{}\ fully in terms of derivatives
\eqn\fuld{\eqalign{
& \la \za_3|
V_{\za_+}(x_5) \,V_{\za_2}(x_2)\big(t L_{-1}^2(x_2)-L_{-2}(x_2)\big)|\za_1\ra
\equiv
\Big(\big(x_2\cdot (\partial_{x_2} +\partial_{x_5})\big)^2\cr
&+{2h\over
t}(1-t) x_2\cdot (\partial_{x_2}
+\partial_{x_5})-{4h\over t}\,\triangle(\za_2) -{2h\over
t} \partial_{x_5} \cdot x_{52} \,
{x_2\cdot x_5\over r_5} \Big)\CA\,.
}}
Dropping the   derivative terms with respect to
$x_5$  gives for the null vector equation
\eqn\fulda{\eqalign{
&\Big((w_2\partial_{w_2})^2 +{2h\over t}(1-t) w_2\partial_{w_2}
-{4h\over t} \triangle(\za_2)  \Big) \int \CA=0\,, \cr
&\Longrightarrow \ \  2a\,({2h\over t} (1-t) +2a) -{4h\over
t}\triangle(\za_2)=0\,.
}}
This  equation for $a$ does not depend on the charge  $\alpha_2$ itself
but rather on the scaling dimension $2\triangle(\za_2)$  and is the
same as the one  for the 3-point matrix element without a
screening charge. Accordingly we obtain as in the
one-dimensional case  $2h=1$ two solutions for $\triangle(\za_3)=a
+\triangle(\za_1)+\triangle(\za_2) $,
\eqn\tpso{
 \triangle(\za_3)=\triangle(\za_2+\za_1)\,, \qquad  \triangle(\za_3)=
\triangle(\za_2-\za_1) =\triangle(\za_2+\za_1+\za_5)\,;
}
the first corresponds to the screeningless case, the second to
the matrix element \tps.

\subsec{The 4-point  null vector decoupling condition}

 We shall now apply the relations \dinv{}\ for the 4-point matrix
element with one screening charge. In this case we can specialise
the argument of the generators $L_{-n}(z_0)$ in \dinv{}\ to  the
coordinate of each of the two middle vertex operators $z_0=x_2$,
or $z_0=x_3$ and thus obtain   two identities
\eqna\nvo
$$\eqalignno{
&0={2h\over t} \int {\rm d}^{2h}x_5
\la \za_{4} |V_{\za_+}(x_5) \, V_{\za_3}(x_3)\, V_{\za_2}(x_2)\,
\big(t L_{-1}^2(x_i)-L_{-2}(x_i)\big)
|\alpha_1\ra \cr
& \equiv \CD_i \int  {\rm d}^{2h} x_5\   \CA
   +\int  {\rm d}^{2h}x_5\ I_i\,, \ \ \  i=2,3\,.  & \nvo {} \cr
}$$
 Here $\CA$ is the matrix element in \exac{}\ and $ \CD_i$ are
differential operators
\eqna\difsing
$$\eqalignno{
 \CD_2&=\CD(x_2,x_3; \za_3)\cr
&=
 (x_2\cdot D)^2 -x_2\cdot D
 -(x_2\cdot\partial_{x_2})^2-\zr^2 (x_3\cdot\partial_{x_3})^2
-2\zr\,\cos\theta\,x_2\cdot\partial_{x_2} x_3\cdot\partial_{x_3}\cr
&+(2h-4)(x_2\cdot D-x_2\cdot \partial_{x_2} -\rho\cos\theta
x_3\cdot \partial_{x_3}) +x_2\cdot \partial_{x_2} +\rho^2
x_3\cdot \partial_{x_3} & \difsing {} \cr
&+\big(2+{2h\over t}(1-t)\big)   \big((1+\rho\cos\theta)(x_2\cdot
D-x_2\cdot \partial_{x_2})
-(\rho^2+\rho\cos\theta)  x_3\cdot\partial_{x_3}\big) \cr
& + 4\ta\tb\, x\,\zr^2\sin^2\theta +{4h\over t}
\triangle(\za_3)\zr^2\sin^2\theta\,, \cr
\CD_3&=\CD(x_3,x_2; \za_2)\,,
}$$
with
$$D=\partial_{x_2}+\partial_{x_3}\,, \ \
\rho^2={x^2_2\over x^2_3}={y\over x}\,, \  \
 2\rho\cos\theta=2{x_2\cdot x_3\over x_3^2}={x+y-1\over x}\,.
$$
Furthermore these operators are expressed as
\eqn\difsingb{
{x_i^2\over x_{23}^2}\,\CD_i=(x_3^2)^{-\Delta}\,
\Big((x+y-1)^2-4xy\Big)\,D_i(x,y)\,
(x_3^2)^{\Delta} \,, \quad i=2,3\,.
}
The  operators $D_i(x,y)$ here
 are given by formulae analogous to
the two  differential operators in \part{}, with the parameters
$\za, \zb, \zg, \zg'$  in \grk{}\ replaced by
 $\ta, \tb, \tg, \tg'$
\eqn\newpar{\eqalign{
 \ta=&
-2\alpha_2\alpha_3+
h+{2h\over
t}(1-t)-2\Delta
\,, \quad \tb=2\za_2\za_3\,, 
\cr
 \tg=& 1+{h\over
t}(1-t)-2 \Delta\,,
\quad  \  \tgg=
1+{h\over t}(1-t)\,,\cr
}}
plus the additional terms
\eqn\addi{
-{h\over t}\,{d(\za_3)\over x}:=
{\Delta(\Delta-{h\over
t}(1-t)) - {h\over
t}\triangle(\alpha_3)\over x}\,,
\qquad \  -{h\over t} {\triangle(\alpha_2) \over y}
}
respectively.
These operators are  precisely the AK operators \part{}\
when the latter are rewritten on the matrix element 
$(x_3^2)^{\Delta}\int \CA$,
which according to \exac{}\ differs by a prefactor from F(x,y).
The asymmetry in \addi\ is due to the choice
of $x_3^2$ as a third independent variable in both equations \difsingb.
\medskip
The integrands $I_i$ of the   remaining integrals in \nvo{}\ are expected
to be expressible as full derivatives in the integration variable
so that these integrals
 vanish identically. Indeed we have checked  this for  the
linear  combination
\eqna\fulld
$$\eqalignno{
&\int  {\rm d}^{2h}x_5 \big(I_1- {r_2\over r_3} I_2\big)= {2h\over t}\int {\rm
d}^{2h}x_5 \Big(
(\partial_{x_5}\cdot x_2\, {x_3\cdot x_5\over r_5} -
\partial_{x_5}\cdot x_3\, {x_2\cdot  x_5\over r_5})\, {x_{52}\cdot
x_{53}\over r_3}\cr
& &\fulld {}\cr
&+ \partial_{x_5}\cdot x_{52}\, (
{x_3\cdot x_5\over r_5}\, {x_2\cdot x_3\over r_3} - {x_2\cdot x_5\over r_5})
 - \partial_{x_5}\cdot
x_{53}\, ({x_2\cdot x_5\over r_5}\, {x_2\cdot x_3\over r_3} -
{x_3\cdot x_5\over r_5} \, {r_2\over r_3}) \Big) {\cal A} = 0\,.
}$$
\medskip
In the screeningless case one recovers the same operators
\difsing{}\ but with different value $\Delta\to $ $
\Delta'= - 2\za_1\za_2-2\za_1\za_3-2\za_2\za_3 $ to be inserted  in \difsingb.
Changing back variables, this correlation function corresponds to a
constant factor $F(x,y)$ with parameters $\za'\, \zb'=0$, so that
it trivially satisfies the initial equations \part{}.

\newsec{Conclusions}

We have revealed a hidden Virasoro symmetry in a 2h-dimensional
model and have demonstrated that it leads to  differential
equations for the 4-point correlation functions in a way analogous to the
2-dimensional case.  This symmetry allows to determine  the
leading short distance behaviour  purely
algebraically, without having to perform the complicated multiple
Mellin integral computation of the Symanzik method.

The Visaroro algebra is related to a dimension
two operator.  One can extend it 
by 
fields  of higher integer dimension, like the
energy-momentum tensor,  which also depends on the
parameter $\alpha_0$ - this problem is left for future investigation; 
see also \NSTb\ for a different approach to an analogous problem.

The four-dimensional model considered here is still an unrealistic,
nonunitary toy model, see also the discussion in \FP. It remains
to be seen whether it could be used as a building block in more
realistic  applications.

\bigskip
\noindent
{\bf  Acknowledgments}
\smallskip

\noindent We thank N. Nikolov, G. Sotkov and I. Todorov for the
interest in this work and for a useful discussion.  P.F.
acknowledges the support of the Italian Ministry of Education,
University and Research (MIUR).  V.B.P.  acknowledges the
hospitality  of INFN, Trieste, and  the University of
Northumbria, Newcastle, UK. This research is supported in part by
the TMR Network EUCLID, contract HPRN-CT-2002-00325, and by 
 the Bulgarian National Council for Scientific Research, grant
F-1205/02.

\medskip
\noindent
\listrefs
\bye